\documentclass[%
aip,
amsmath,amssymb,
reprint,%
]{revtex4-2}
\usepackage{graphicx}
\usepackage{dcolumn}
\usepackage{mathrsfs}
\usepackage{bm}
\usepackage{lineno}
\usepackage{hyperref}
\usepackage{bbm}
\usepackage{ulem}
\usepackage{xcolor}

\begin{document}

\title{Temporal coherence of optical fields in the presence of entanglement}

\author{Yunxiao Zhang}
 \affiliation{College of Precision Instrument and Opto-Electronics Engineering, Key Laboratory of Opto-Electronics Information Technology, Ministry of Education, Tianjin University, Tianjin 300072, P. R. China}

\author{Nan Huo}
 \affiliation{College of Precision Instrument and Opto-Electronics Engineering, Key Laboratory of Opto-Electronics Information Technology, Ministry of Education, Tianjin University, Tianjin 300072, P. R. China}

\author{Liang Cui}
 \affiliation{College of Precision Instrument and Opto-Electronics Engineering, Key Laboratory of Opto-Electronics Information Technology, Ministry of Education, Tianjin University, Tianjin 300072, P. R. China}

\author{Xueshi Guo}%
\affiliation{College of Precision Instrument and Opto-Electronics Engineering, Key Laboratory of Opto-Electronics Information Technology, Ministry of Education, Tianjin University, Tianjin 300072, P. R. China}

\author{Jiahao Fan}
\affiliation{Department of Physics, City University of Hong Kong, 83 Tat Chee Avenue, Kowloon, Hong Kong, P. R. China}

\author{Zhedong Zhang}
\affiliation{Department of Physics, City University of Hong Kong, 83 Tat Chee Avenue, Kowloon, Hong Kong, P. R. China}

\author{Xiaoying Li}
 \email{xiaoyingli@tju.edu.cn}
\affiliation{College of Precision Instrument and Opto-Electronics Engineering, Key Laboratory of Opto-Electronics Information Technology, Ministry of Education, Tianjin University, Tianjin 300072, P. R. China}%

\author{Z. Y. Ou}
 \email{jeffou@cityu.edu.hk}
\affiliation{Department of Physics, City University of Hong Kong, 83 Tat Chee Avenue, Kowloon, Hong Kong, P. R. China}

\begin{abstract}

In classical coherence theory, coherence time is typically related to the bandwidth of the optical field. Narrowing the bandwidth will result in the lengthening of the coherence time. This will erase temporal distinguishability of photons due to time delay in pulsed photon interference. However, this is changed in an SU(1,1)-type quantum interferometer where quantum entanglement is involved. In this paper, we investigate how the temporal coherence of the fields in a pulse-pumped SU(1,1) interferometer changes with the bandwidth of optical filtering. We find that, because of the quantum entanglement, the coherence of the fields does not improve when optical filtering is applied, in contrary to the classical coherence theory, and quantum entanglement plays a crucial role in quantum interference in addition to distinguishability.

\end{abstract}

\maketitle

In quantum mechanics, indistinguishability is a fundamental concept for quantum interference.  When there exist distinguishable paths for two interfering fields, no interference can occur. This is at the heart of Bohr's complementarity principle for quantum interference \cite{bohr} and is summarized quantitatively in a relation as  \cite{dis,remp}
\begin{eqnarray}\label{VD}
{\cal V}^2+{\cal D}^2\le 1
\end{eqnarray}
between the visibility of interference ${\cal V}$ and the path distinguishability ${\cal D}$. On the other hand, path distinguishability can be erased by projection measurement in quantum erasers \cite{scully1,scully2}.
This erasure of distinguishability can also be achieved in the measurement process. For example, optical filtering can increase the coherence time of an optical field. In an interference experiment involving optical pulses, when optical delay is introduced to achieve distinguishability in time, no interference occurs by direct detection. But placing a narrow band filter in front of the detector can erase the temporal distinguishability by lengthening the pulse and recover the interference effect. This is what leads to the important concept of coherence time in classical coherence theory. It is the time interval in which an optical field keeps its phase correlation and according to the classical coherence theory, it is typically related to the reciprocal bandwidth of an optical field \cite{bw} and can be lengthened by optical filtering. Indeed, a Gaussian-shaped field passing through a filter of Gaussian profile of width $\sigma_f$ has its coherence time changed to (see Supplementary Materials I for details)
\begin{eqnarray}\label{Tc}
T_c' = \sqrt{T_c^2 + 1/\sigma_f^2}
\end{eqnarray}
with $T_c$ as its original coherence time.

\begin{figure}[t]
	\includegraphics[width=8cm]{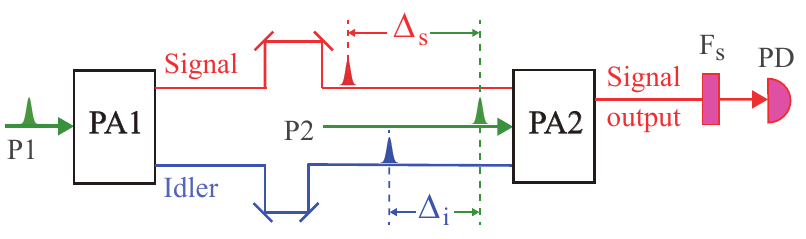}
	\caption{An unbalanced SU(1,1) interferometer (SUI) with filtered detection. PA, parametric amplifier; P, pump; $F_s$ narrow band filter; PD, power detector.}
	\label{SUI}
\end{figure}

On the other hand, quantum entanglement exists for quantum fields, which was recently shown \cite{qian16,qian18}  to be the missing third quantity for Eq.(\ref{VD}) to take the equal sign. With entanglement, Eq.(\ref{VD}) takes the new form of
\begin{eqnarray}\label{VDC}
{\cal V}^2+{\cal D}^2+{\cal C}^2 =  1,
\end{eqnarray}
where quantity ${\cal C}$ is the concurrence that is a measure of entanglement.  The strong quantum correlations between two entangled fields give ${\cal C} =1$ and can thus lead to disappearance of coherence (${\cal V}=0$) even if there is indistinguishability (${\cal D}=0$) \cite{qian19}. Moreover, quantum entanglement makes it possible to manipulate the coherence property of one field by controlling the other entangled field \cite{zou91,zaj,zei19}. All this poses a challenge for the aforementioned traditional concept of temporal coherence as related to the reciprocal bandwidth of the field: narrow filtering leads to the temporal indistinguishability but does not deal with entanglement, which can reduce interference effect as well according to Eq.(\ref{VDC}).

An SU(1,1) interferometer is a quantum interferometer that is a perfect platform for this study. It utilizes parametric amplifiers (PA) to replace beam splitters in traditional classical interferometers \cite{yur,che}. The signal and idler fields generated in the parametric amplifier are entangled to each other and are used to probe phase changes in the interferometer. Such a novel quantum interferometer has recently been widely applied to quantum imaging \cite{lem14}, quantum sensing \cite{kri15,kri20,tera,che15}, quantum state engineering \cite{su19,cui20}, and quantum measurement \cite{nc,JMLI-OE19,JMLI-PRA20}. Because of the involvement of quantum entanglement in the interferometer, its properties can be quite different from traditional classical interferometer \cite{ou-li20}, especially when it concerns the temporal coherence of the fields involved.

\begin{figure}[t]
	\includegraphics[width=7.0cm]{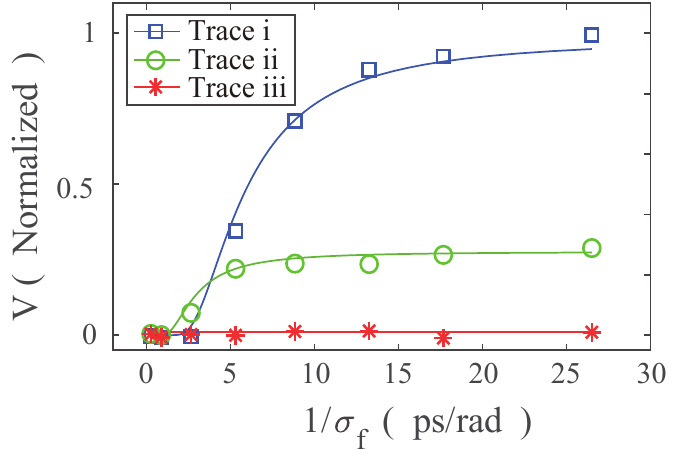}
	\caption{Visibility as a function of reciprocal bandwidth of the filter. (i) $\Delta_i=0, \Delta_s =10 ps$; (ii) $\Delta_i= 5 ps, \Delta_s = 10 ps$; (iii) $\Delta_i= 10 ps, \Delta_s = 10 ps$.}
	\label{filter}
\end{figure}

For studying this, we consider a pulse-pumped SU(1,1) interferometer shown in Fig.\ref{SUI}. The pumps of two PAs, P1 and P2, are originated from the same laser. When the paths are carefully balanced for all the fields involved, interference is observed in both outputs. Because of symmetry, we only need to consider detection at one output, say, the signal port. We now introduce  delays ($\Delta_s, \Delta_i$) in either of the arms so that interference disappears due to temporal distinguishability. We next place narrow optical filters at the signal output port in front of the detector with the intention to lengthen the coherence time and erase the temporal distinguishability.   This works for the delay in signal field only ($\Delta_s \ne 0$ but $\Delta_i =0$), as shown in the blue data points (Trace i) in Fig.\ref{filter}, where the visibility increases with the narrowing of the filter and eventually reaches 100\%. But it fails if there is a delay in the idler field ($\Delta_i \ne 0$), as shown in the green (Trace ii) and red (Trace iii) data points in Fig.\ref{filter} (see Supplementary Materials IV-A and IV-B for details of measurement). Notice especially the large $\Delta_i$ delay case (Trace iii) with nearly zero visibility no matter how narrow the filter is. This is in complete contradiction with classical coherence theory and is an example of the influence of quantum entanglement on coherence, as we will see in the following.

Although the increase of the visibility in Fig.\ref{filter} can be attributed to narrowing of the filter as classical coherence theory, the limiting values of $1/\sigma_f\rightarrow \infty$ can only be understood with the language of Eq.(\ref{VDC}) by a single-mode two-photon state in the form of
\begin{eqnarray}\label{st}
|\Psi\rangle =A_1|s_1,i_1\rangle + A_2|s_2,i_2\rangle,
\end{eqnarray}
where $A_1, A_2$ are related to the pump fields of PA1 and PA2, respectively and are normalized: $|A_1|^2+| A_2|^2=1$. $s_1, i_1$ and $s_2,i_2$ are the signal and idler modes from two PAs, respectively. Normally, they are distinct and independent. When we align the two idler fields together, depending on the delay in the idler field, the two idler modes will change from totally independent when $\Delta_i$ is much larger than the pulse width to completely identical when $\Delta_i=0$.

For the latter case when $\Delta_i=0$, the idler photon states from two PAs are identical: $|i_1\rangle = |i_2\rangle \equiv |i\rangle$ so Eq.(\ref{st}) becomes
\begin{eqnarray}\label{st2}
|\Psi\rangle(\Delta_i=0) = (A_1|s_1\rangle + A_2|s_2\rangle)\otimes |i\rangle,
\end{eqnarray}
which is a product state with no entanglement or concurrence ${\cal C}=0$. Then interference observed at the signal output port is related to the indistinguishability in the signal field between $|s_1\rangle$ and $|s_2\rangle$, which can be altered by the optical filtering of the signal field. This corresponds to the classical case when Eq.(\ref{VD}) takes equal sign and is described by the blue data in Fig.\ref{filter}. The solid blue curve is a normalized Gaussian function with a width given by Eq.(\ref{Tc}) (see later for detail).

In the other extreme case when there is a large delay in idler field, $|i_1\rangle$ and $|i_2\rangle$ are well separated and the state in Eq.(\ref{st}) becomes a time-bin entangled state with maximum entanglement for $A_1=A_2=1/\sqrt{2}$, leading to ${\cal C}=1$ but this also gives ${\cal D}=0$ according to Ref.\citenum{qian19}. According to Eq.(\ref{VDC}), this should give ${\cal V} = 0$, which is indeed what we observed as the red points in Fig.\ref{filter}.

The intermediate case for the green data in Fig.\ref{filter} can be understood by evaluating concurrence ${\cal C}$ for the state in Eq.(\ref{st}) with $|i_2\rangle$ partially overlapping with $|i_1\rangle$: $|i_2\rangle = \cos \theta |i_1\rangle + \sin\theta |u\rangle$, where  $\cos \theta = \langle i_1|i_2\rangle $ and $|u\rangle$ is some state orthonormal to $|i_1\rangle$ and can be obtained from Schmidt method.  Working in the space spanned by $\{|s_1\rangle,|s_2\rangle,|i_1\rangle,|u\rangle\}$, we find the concurrence ${\cal C}=2|A_1A_2|\sin \theta$ for the state in Eq.(\ref{st}) (See Supplementary Materials II). It is straightforward to calculate the visibility for interference between fields $s_1,s_2$, yielding ${\cal V} = 2|A_1A_2|\cos \theta$ and according to Ref.\citenum{qian19}, the distinguishability quantity ${\cal D} = |(|A_1|^2-|A_2|^2)|$. It can be checked that these values of ${\cal V}, {\cal C}, {\cal D}$ satisfy Eq.(\ref{VDC}), confirming the role played by entanglement in interference.

One may attempt to reach a full understanding results in Fig.\ref{filter} by Eq.(\ref{VDC}) but this is very complicated because it involves concurrence of a multi-mode two-photon state in a continuous temporal/spectral space. Nevertheless, we can understand the results by calculating the visibility of interference from the multi-mode two-photon state:
\begin{eqnarray}\label{Phi}
|\Phi_2\rangle &=& \int d\omega_1 d\omega_2 \Big[ \Phi(\omega_1,\omega_2) e^{i\omega_1\Delta_s}e^{i\omega_2\Delta_i} + \Phi'(\omega_1,\omega_2) \Big]\cr
&&\hskip 0.5in \times\hat a_s^{\dag}(\omega_1)\hat a_i^{\dag}(\omega_2)|vac\rangle
\end{eqnarray}
where $\Phi(\omega_1,\omega_2), \Phi'(\omega_1,\omega_2)$ are the two-photon wave function for the two-photon states produced by PA1 and PA2 respectively and take the form of $A_p^2 \exp[-(\omega_1+\omega_2-2\omega_{p0})^2/4\sigma_p^2] F(\omega_1,\omega_2)$ for a four-wave mixing process \cite{chen05} with a Gaussian-shaped pump spectral profile (peak amplitude $A_p$, width $\sigma_p$, center frequency $\omega_{p0}$) and $F(\omega_1,\omega_2)$ as the phase matching factor depending on nonlinear media.
The field at signal output after the filter ($F_s$) but before the detector is
\begin{eqnarray}\label{E-f}
\hat E_s(t) = \int d\omega \hat a_s(\omega)e^{-i\omega t} f_s(\omega)
\end{eqnarray}
where $f_s(\omega)$ is the filter amplitude transmission function.

Assume the detector is slow and cannot trace the profile of single pulses. Then the output of the detector is
\begin{eqnarray}\label{ifout-i}
i_D
&\propto & \int d\tau \langle \hat E_s^{\dag}(\tau)\hat E_s(\tau)\rangle\cr
&\propto &  \int d\omega_1d\omega_2 | f_s(\omega_1)|^2 |\Phi_{eq}(\omega_1,\omega_2)|^2
\end{eqnarray}
with $\Phi_{eq}(\omega_1,\omega_2)\equiv \Phi(\omega_1,\omega_2) e^{i\omega_1\Delta_s+i\omega_2\Delta_i} + \Phi'(\omega_1,\omega_2)$.

Assume the two PAs are identical and there is a delay of $\Delta_p$ for the pump of PA2 with respect to PA1. Then,
$\Phi'(\omega_1,\omega_2)$  $=\Phi(\omega_1,\omega_2)e^{i(\omega_1+\omega_2)\Delta_p} $ and $|\Phi_{eq}(\omega_1,\omega_2)|^2= |\Phi(\omega_1,\omega_2)|^2|e^{i\omega_1(\Delta_s-\Delta_p)}e^{i\omega_2(\Delta_i-\Delta_p)}+1|^2 $. So, if we treat $\Delta_s, \Delta_i$ as delays reference to the pump of PA2, we can drop $\Delta_p$. Then we have from Eq.(\ref{ifout-i})
\begin{eqnarray}\label{ifout-i2}
i_D
&\propto& \int d\omega_1d\omega_2 | f_s(\omega_1)|^2 |\Phi(\omega_1,\omega_2)|^2 |e^{i\omega_1\Delta_s}e^{i\omega_2\Delta_i}+1|^2\cr
&\propto & 1+{\cal V}(\Delta_s,\Delta_i) \cos(\omega_{10}\Delta_s+\omega_{20}\Delta_i),
\end{eqnarray}
where $\omega_{10}$ is the center frequency of the filter for the signal field and with $\omega_{p0}$ as the center frequency of the pump field, $\omega_{20} = \omega_{p0}-\omega_{10}$ is the center frequency of the idler field. So, interference is exhibited with visibility
\begin{eqnarray}\label{V0}
&&{\cal V}(\Delta_s,\Delta_i) \cr &&\hskip 0.2in \equiv \frac{\Big|\int d\omega_1d\omega_2 | f_s(\omega_{1})|^2 |\Phi(\omega_{1},\omega_{2})|^2 e^{i\omega_1\Delta_s}e^{i\omega_2\Delta_i}\Big|}{\int d\omega_1d\omega_2 | f_s(\omega_1)|^2 |\Phi(\omega_1,\omega_2)|^2 }.~~~~~~~
\end{eqnarray}

Let us assume a Gaussian shape for $f_s(\omega_1)$:
\begin{eqnarray}\label{fi}
f_s(\omega_1) = \exp[-(\omega_1-\omega_{10})^2/2\sigma_f^2]
\end{eqnarray}
with a width of $\sigma_f$.  We can assume for simplicity the phase match factor in  $\Phi(\omega_1,\omega_2)$ also has a Gaussian shape as $F(\omega_1,\omega_2)=\exp[-(\omega_1-\omega_{2})^2/2\sigma_0^2]$ with a width of $\sigma_0$. Then, after shifting to center frequencies of all fields with $\Omega_j\equiv \omega_j-\omega_{j0}$ ($j=1,2$), we have
\begin{eqnarray}\label{fi}
|f_s(\omega_1)\Phi(\omega_1,\omega_2)|^2 &=& \exp\bigg(-\frac{\Omega_1^2}{\sigma_f^2}\bigg)
\exp\bigg[-\frac{(\Omega_1+\Omega_2)^2}{2\sigma_p^2}\bigg]\cr
&&\hskip 0.3in \times\exp\bigg[-\frac{(\Omega_1-\Omega_2)^2}{\sigma_0^2}\bigg].~~~~
\end{eqnarray}
From Eq.(\ref{V0}), we obtain
\begin{eqnarray}\label{V}
{\cal V}(\Delta_s,\Delta_i) = \exp[- Q(\Delta_s,\Delta_i)/4],
\end{eqnarray}
with
\begin{eqnarray}\label{Q}
Q(\Delta_s,\Delta_i) &\equiv & \frac{2\Delta_i^2\sigma_p^2\sigma_0^2}{2\sigma_p^2+\sigma_0^2} +\Big(\Delta_s-\frac{\sigma_0^2-2\sigma_p^2}{\sigma_0^2+2\sigma_p^2}\Delta_i\Big)^2\cr &&\hskip 0.7in \times \frac{\sigma_0^2+2\sigma_p^2}{4+(\sigma_0^2+2\sigma_p^2)/\sigma_f^2}.
\end{eqnarray}

The visibility expression in Eq.(\ref{V}) can explain the results in Fig.\ref{filter}.
For Trace (i) in Fig.\ref{filter}, the idler path is balanced with $\Delta_i=0$, but signal field has delay $\Delta_s$. Then we have
\begin{eqnarray}\label{V1}
{\cal V}_1(\Delta_s) = \exp[-\Delta_s^2/4T_s^2],
\end{eqnarray}
with
\begin{eqnarray}\label{Ts}
T_s^2\equiv \frac{1}{\sigma_f^2}+\frac{4}{\sigma_0^2+2\sigma_p^2}.
\end{eqnarray}
The solid curve for Trace (i) of Fig.\ref{filter} is a plot of Eq.(\ref{V1}) as a function of $1/\sigma_f$ with $\sigma_p=0.28$ rad/ps, $\sigma_0= 4.8$ rad/ps taken from experimental parameters (see Supplementary Materials IV-A).  Note that Eq.(\ref{Ts}) is actually the coherence time of the signal field after the filter. To see better the physical meaning of Eq.(\ref{Ts}), we rewrite it in terms of $T_f\equiv 1/\sigma_f$, $T_0\equiv 1/\sigma_0$, and the coherence time $T_p \equiv 1/\sigma_p$ of the pump field:
\begin{eqnarray}\label{Ts2}
T_s^2 = T_f^2+T_c^2
\end{eqnarray}
with
$T_c^2\equiv 4T_p^2T_0^2/(2T_0^2+T_p^2)$ as the coherence time of the generated signal field
without filter. Note that Eq.(\ref{Ts2}) has the same form as Eq.(\ref{Tc}) for the classical interferometer so $T_s$ depends highly on  $T_f$, the reciprocal width of the filter, and is dominated by $T_f$ for narrow filter width $\sigma_f\rightarrow 0$ but reaches $T_c$ for no filtering. This indicates that optical filtering of the signal field can increase the coherence time of the signal field, which is the same as the classical interferometer. This is not surprising because $\Delta_i=0$ corresponding to the case in Eq.(\ref{st2}) as discussed earlier and ${\cal C}=0$ or Eq.(\ref{VDC}) becomes Eq.(\ref{VD}) with equal sign for classical case.

However, as $\Delta_i$ becomes non-zero and is large enough so that $|i_1\rangle$ and $|i_2\rangle$ are well separated, the visibility is nearly zero independent of filtering, as shown in the red points of Fig.\ref{filter} and the solid curve of Trace (iii) obtained from Eq.(\ref{V}) with $\Delta_i = 10 ps$ but $\sigma_p, \sigma_0$ kept the same as before. The intermediate case of green points and Trace (ii) is for $\Delta_i = 5 ps$. It can be seen that the theoretical prediction from Eq.(\ref{V}) fits experimental data very well.

As another test of Eq.(\ref{VDC}), we consider an extreme case of ${\cal D} =0$ by setting
 $\Delta_s=0$ with balanced signal path so that signal photon wavepackets overlap in time and become completely indistinguishable in its generation from either PA1 or PA2. One may think filtering at signal field would have no effect on interference since we already have complete indistinguishability for the signal photon.  However, because of the quantum correlation between the signal and idler photons, filtering at signal field may alter the idler field and according to the discussion on the state in Eq.(\ref{st}), this will change
the entanglement property between the signal and idler fields, which, in other words, is concurrence ${\cal C}$. This will change interference visibility ${\cal V}$ according to Eq.(\ref{VDC}), as we will see in the following.

When $\Delta_s=0$,  we have from Eq.(\ref{Q})
\begin{eqnarray}\label{V2}
{\cal V}_2(\Delta_i) = \exp[-\Delta_i^2/4T_i^2]
\end{eqnarray}
with
\begin{eqnarray}\label{Ti}
T_i^2&\equiv &\frac{\sigma_0^2+2\sigma_p^2+4\sigma_f^2}{\sigma_f^2(\sigma_0^2+2\sigma_p^2)+2\sigma_p^2\sigma_0^2}\cr
&=&  \frac{2T_f^2T_0^2+4T_0^2T_p^2+T_f^2T_p^2}{2T_f^2+2T_0^2+T_p^2}.
\end{eqnarray}
Now let us try to understand Eq.(\ref{Ti}) in the language of Eq.(\ref{VDC}). As said earlier, $\Delta_s=0$ corresponds to ${\cal D}=0$ and Eq.(\ref{VDC}) becomes
\begin{eqnarray}\label{VC}
{\cal V}^2+{\cal C}^2 = 1.
\end{eqnarray}

Now ${\cal C}$ plays a similar role as ${\cal D}$ in Eq.(\ref{VD}) with equal sign. In this case, ${\cal C}$ can be altered by delay $\Delta_i$ of the idler field, as
we discussed for Eq.(\ref{st}). Delay $\Delta_i$ controls the overlap $\cos\theta = |\langle i_1|i_2\rangle| $ of the two idler fields and then the degree of entanglement: $\Delta_i =0$ leads to product state with no entanglement while large $\Delta_i $ may give maximum entanglement with concurrence ${\cal C} =1$. This is reflected directly in Eq.(\ref{V2}) and the size of relevant $\Delta_i $ is determined by $T_i$ in Eq.(\ref{Ti}), which depends on the filter bandwidth $\sigma_f$.

There are two notable limiting cases with simple physical pictures.  For narrow band pumping ($\sigma_p \ll \sigma_f, \sigma_0$), $T_i\rightarrow  \sqrt{T_f^2+4T_0^2}$, which is similar to $T_s$ in Eq.(\ref{Ts2}) or classical coherence time in Eq.(\ref{Tc}) and strongly depends on $T_f$ for narrow filtering of the signal field. Perhaps it is  easier to understand this case in the extreme case of CW pumping with $\sigma_p =0$. Under this condition, the frequencies of signal and idler photons are perfectly anti-correlated: $\omega_s+\omega_i = 2\omega_{p0}$ so filtering signal field is equivalent to filtering the idler field and will also narrows the idler bandwidth and therefore changes $T_i$ in the same way. That is why $T_i$ in this case has the same form as $T_s$ in Eq.(\ref{Ts2}) or the classical coherence time in Eq.(\ref{Tc}). Filtering can change ${\cal C}$ and thus ${\cal V}$, in a similar way to ${\cal D}$ in Eq.(\ref{VD}).

On the other hand, for broad band pumping ($\sigma_f \ll \sigma_p, \sigma_0$), $T_i\rightarrow \sqrt{T_p^2/2+T_0^2} \equiv \bar T_i$, which hardly depends on $\sigma_f$, the width of the filter. In this limiting case, the broad band pumping results in imperfect frequency correlation between signal and idler. Thus optical filtering for the signal field does not significantly change the spectral/temporal property of the correlated idler fields or the overlap $\cos\theta$ of the two idler fields, which is mainly determined by the pump bandwidth $\sigma_p$ and phase matching bandwidth $\sigma_0$.

\begin{figure}[t]
	\includegraphics[width=6.5cm]{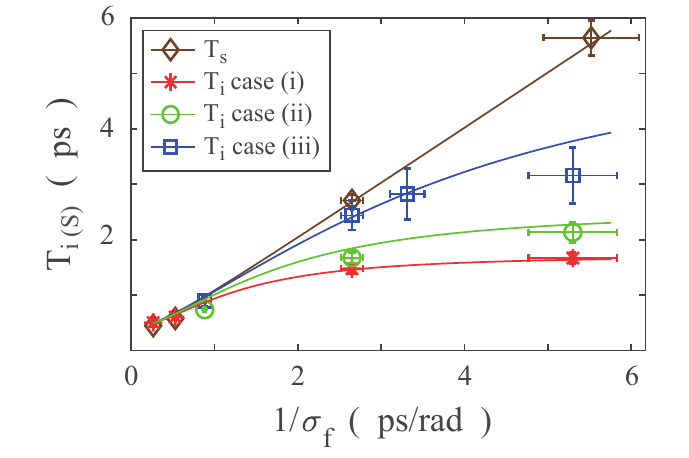}
	\caption{$T_i$ as a function of reciprocal  bandwidth $1/\sigma_f$ of the signal field filter for (i) $\sigma_p = 0.80 nm$, (ii) $\sigma_p= 0.60 nm$, and (iii) $\sigma_p= 0.24 nm$.  $T_s$ is also plotted from Eq.(\ref{Ts}) together with measured data, corresponding to $\sigma_p=0$ for $T_i$.}
	\label{TiTs}
\end{figure}

The intermediate case of finite pumping bandwidth can be fully described by Eq.(\ref{Ti}). To verify it, we measure $T_i$ for three bandwidths of pumping as a function of the filter bandwidth. $T_i$ is obtained  by measuring visibility versus delay $\Delta_i $ and fitting Eq.(\ref{V2}) (see Supplementary Materials IV-B and IV-C for details). The results are shown in Fig.\ref{TiTs} and the experimental data fit the theory quite well. As a comparison, we also plot $T_s$ (brown), which is the same as $T_i$ in the limit of $\sigma_p=0$ and linearly depends on $1/\sigma_f$. As the pump bandwidth increases, $T_i$ deviates from this linear dependence more and more.

In summary, by investigating how optical filtering can affect the performance of an unbalanced pulsed SU(1,1) quantum interferometer, we find that quantum entanglement plays a similar role as distinguishability in changing the visibility of interference.  On the other hand, this connection between interference visibility and quantum entanglement can be used to characterize quantum entanglement from measured visibility for some systems that are hard to access \cite{QE}.

{It should be noted that our discussion about the relation between interference and entanglement is based on a single-mode description of the system. For the multi-mode description of experiment given in Eq.(\ref{Phi}), however, although visibility of interference can be derived and matches the experimental data very well, the simple physical interpretation in the language of Eq.(\ref{VDC}) does not apply because it is a challenge to obtain concurrence ${\cal C}$ for a multi-mode system.

On the other hand, if the joint spectral function $\Phi(\omega_1,\omega_2)$ is factorized so that the fields can be described by single temporal modes, it is equivalent to a single-mode system and can be described in the language of Eq.(\ref{VDC}). This is the case when $\sigma_f\rightarrow 0$ or in the limit of large $1/\sigma_f$ in Fig.\ref{filter}. In this case, we can show (see Supplementary Materials III)
${\cal C}^2=1 - \exp\big(-\Delta_i^2/2\bar T_i^2\big)$. Combining with visibility in Eq.(\ref{Ti}) in the limit of $\sigma_f\rightarrow 0$, we find ${\cal C}$ and ${\cal V}$ satisfy Eq.(\ref{VC}), which is a result of Eq.(\ref{VDC}) when noticing that the limit of large $1/\sigma_f$ means temporal indistinguishability, i.e.,  ${\cal D}=0$, just like the single-mode discussion.
}

~\\
This work was supported in part by National Natural Science Foundation of China (Grants No. 91836302, No. 12004279, and No. 12074283)

\end{document}


\title{Supplementary Materials\\
Temporal coherence of optical fields in the presence of entanglement}

\author{Yunxiao Zhang}

\author{Nan Huo}

\author{Liang Cui}

\author{Xueshi Guo}%

\author{Jiahao Fan}

\author{Zhedong Zhang}

\author{Xiaoying Li}

\author{Z. Y. Ou}

\maketitle

\section{Classical Mach-Zehnder interferometer (MZI) with optical filter}

\begin{figure}[t]
	\includegraphics[width=8cm]{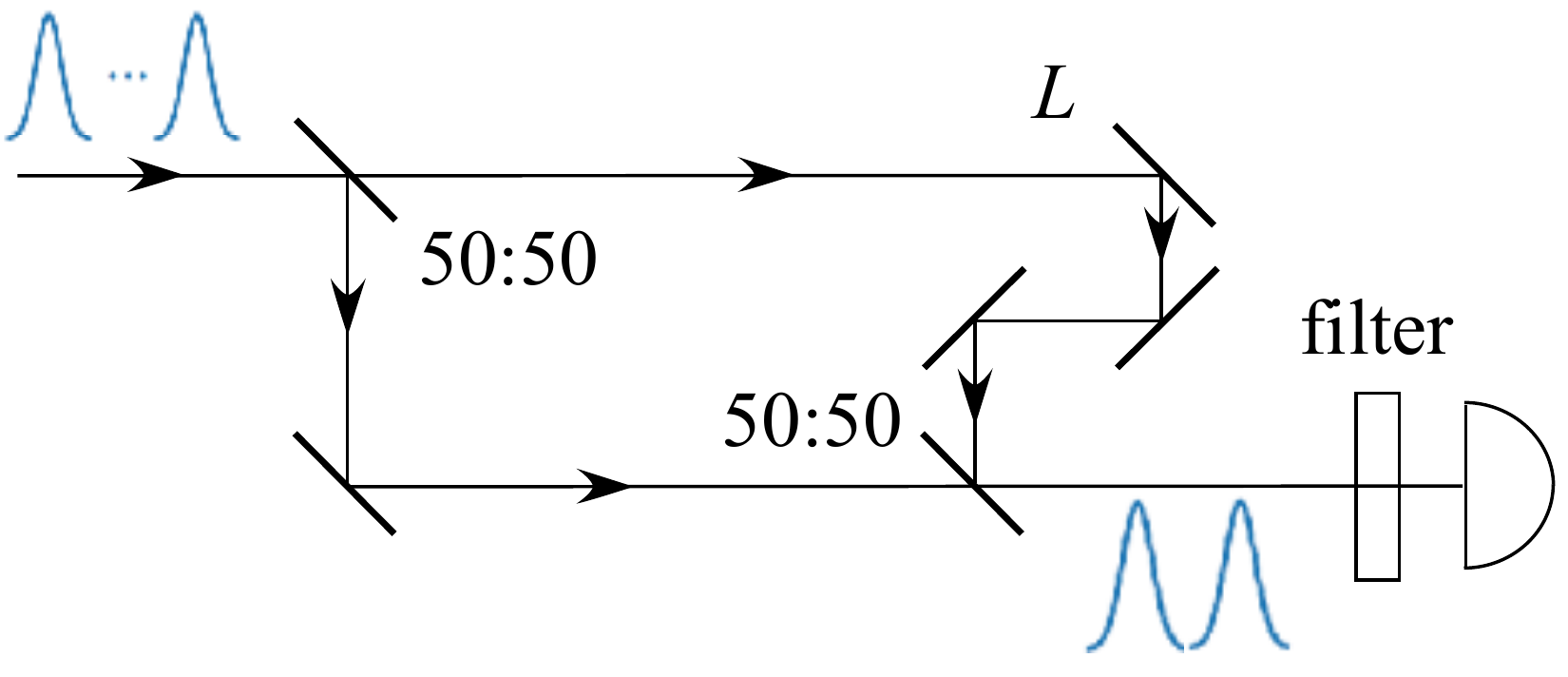}
	\caption{An unbalanced Mach-Zehnder interferometer (MZI) with filtered detection. }
	\label{MZ}
\end{figure}

To compare the results of classical and quantum interferometers, we consider an unbalanced classical Mach-Zehnder interferometer (MZI) for a train of pulses input to the interferometer, as shown in Fig.\ref{MZ}.
The pulse train is described by
\begin{eqnarray}
E(t) = \sum_j  E_{j} (t-j\Delta T)
\end{eqnarray}
with $E_{j} (t) \equiv A_j h(t)$ for the single $j$-th  pulse with amplitude $A_j$ and the same normalized shape $h(t)$ $(\int dt |h(t)|^2 =1)$ for all pulses. The width of $h(t)$ is denoted as $\delta t$.  Denote $h(\omega) \equiv \int dt h(t) e^{-i\omega t}$ as the Fourier component of $h(t)$. With the paths of the two arms unbalanced by a relative delay of $L/c$, the field at the output of the MZI is
\begin{eqnarray}\label{Eout}
E_{out}(t) &\propto & E(t) + E(t+L/c) \cr
&=&  \sum_j  E_{j} (t-j\Delta T) +E_{j} (t+L/c-j\Delta T)
\end{eqnarray}

Now let us place an optical filter of amplitude transmissivity $f(\omega)$ at the output of the interferometer. The field after the filter is then
\begin{eqnarray}\label{Efout}
E_{out}'(t) =  \sum_j  E_{j} '(t-j\Delta T) +E_{j}' (t+L/c-j\Delta T)~~~~
\end{eqnarray}
with $E_{j}' (t) \equiv A_j h'(t)$. $h'(t)$, whose Fourier component is $h'(\omega) =h(\omega)f(\omega)$, is the new pulse shape after the filter with a width of $\delta t'$. For the simplicity of argument, we assume that $\delta t', L/c < \Delta T$, that is, there is no overlap between different pulses in the pulse train even after filtering. With this assumption, it is straightforward to find the intensity after the filter as
\begin{eqnarray}\label{Iout}
I_{out}'(t) &=& \sum_j  |E_{j} '(t-j\Delta T)|^2 +|E_{j}' (t+\frac{L}{c}-j\Delta T) |^2\cr
&&  + \sum_j {E_{j}'}^* (t-j\Delta T) E_{j}' (t+\frac{L}{c}-j\Delta T) +c.c.~~~~
\end{eqnarray}

For a detector with a response function of $k(\tau)$ whose width is the resolution time $T_R$,
the photo-current of the detector for the filtered field is
\begin{eqnarray}\label{iout}
i_D(t) &=& \int d\tau k(t-\tau) I_{out}'(\tau).
\end{eqnarray}
For a slow detector that cannot resolve shape of a single pulse, its time resolution $T_R \gg \delta t'$. Then Eq.(\ref{iout}) can be approximated as
\begin{eqnarray}\label{ioutx}
i_D(t)
&\approx & \sum_j k(t-j\Delta T) |A_j |^2\int d\tau \Big [|h'(\tau)|^2 \cr
&& \hskip 0.3in  + | h'(\tau+L/c)|^2+   { h'}^*(\tau) h'(\tau+L/c) +c.c. \Big ]\cr
&=& 2 Q(t) (1 + {\cal V}_M \cos \Delta \varphi),
\end{eqnarray}
with $Q(t) \equiv \sum_j k(t-j\Delta T) |A_j |^2 \int d\tau |h'(\tau)|^2$, visibility ${\cal V}_M(L/c)\equiv  |\int d\tau {h'}^*(\tau) h'(\tau+L/c)|$ $ /\int d\tau |h'(\tau)|^2$ and phase $\Delta \varphi \equiv {\rm Arg}[\int d\tau {h'}^*(\tau) h'(\tau+L/c)]$.

Take a Gaussian shape for both the initial pulse and the filter: $h(\omega) = e^{-(\omega-\omega_0)^2/2\sigma^2}$, $f(\omega) =e^{-(\omega-\omega_0)^2/2\sigma_f^2}$, which gives $h(t)$ $=e^{-t^2/2\delta t^2}$, $h'(t)$ $=e^{-t^2/2{\delta t'}^2}$ with $\delta t=$ $1/\sigma$ and ${\delta t'}=\sqrt{1/\sigma^2+1/\sigma_f^2}$ , we find
${\cal V}_M(\Delta)=$ $e^{-\Delta^2/4{T_c'}^2}$  $(\Delta \equiv L/c)$  with
\begin{eqnarray}\label{Tc2}
{T_c'}^2 \equiv {T_c'}^2 + T_f^2 = {\delta t'}^2,
\end{eqnarray}
where $ T_f\equiv 1/\sigma_f$. Here, $\delta t={T_c}$ and $\delta t' ={T_c'}$ are the pulse widths or coherence time of the initial and filtered pulses, respectively. So, if the delay $\Delta =L/c \gg \delta t$, no interference occurs without optical filtering ($\sigma_f=\infty$ and $T_f=1/\sigma_f=0$). But the interference can be recovered  with ${\cal V} \approx 100\%$ if the filter is narrow enough so that $T_c'$ is much larger than the delay $\Delta =L/c$. The filter lengthens the coherence time of the pulses so that they will overlap at the BS and interfere. This is consistent with the coherence time concept in classical coherence theory.

\section{The calculation of concurrence for entangled photon pairs}

The states of entangled photon pairs produced from the experiment in main text is in general of the form
\begin{equation}\label{Psi}
   |\Psi\rangle=e^{i \phi}\sqrt{f}|s_1\rangle\otimes |i_1\rangle+\sqrt{1-f}|s_2\rangle\otimes|i_2\rangle,
\end{equation}
where $\sqrt{f}\equiv |A_1|$ and $\sqrt{1-f}\equiv |A_2|$ are related to the pump fields of PA1 and PA2, respectively. Since $\langle {s_1}|{s_2}\rangle= 0$ and  $\langle{i_1}|{i_2}\rangle\neq 0$ (with $0\leq f\leq 1$, $0\leq \phi \leq 2 \pi$) , state $|s_1\rangle$ is found to be orthonormal to $|s_2\rangle$, while states $|i_1\rangle$ and $|i_2\rangle$ are not. In order to find a certain state  $|u\rangle$ which is orthonormal to state $|i_1\rangle$, one can make decomposition of the state $|i_2\rangle=\cos{\theta} |i_1\rangle+\sin{\theta} |u\rangle$, with $0\leq \theta \leq \frac{\pi}{2}$. Note that any extra phase between $|i_1\rangle$ and $|i_2\rangle$ can be absorbed in $e^{i\phi}$ and in state $|u\rangle$. The state in Eq.(\ref{Psi})  thus reads
\begin{equation}
\begin{aligned}
 |\Psi\rangle = & e^{i \phi}\sqrt{f}|s_1\rangle\otimes |i_1\rangle+\sqrt{1-f}\cos{\theta}|{s_2}\rangle\otimes |i_1\rangle\\
   &+\sqrt{1-f}\sin{\theta}|s_2\rangle\otimes|u\rangle.
\end{aligned}
\end{equation}

Having the entangled state in the new orthonormal basis ({$|s_1\rangle\otimes|i_1\rangle, |s_2\rangle\otimes|i_1\rangle, |s_2\rangle\otimes|u\rangle, |s_1\rangle\otimes|u\rangle$}), and defining the Pauli matrices
\begin{equation*}
\begin{split}
    &\sigma_z^s=|s_2\rangle\langle s_2|-|s_1\rangle \langle{s_1}|,\\
    &\sigma_z^i=|i_1\rangle\langle {i_1}|-|u\rangle\langle {u}|.
\end{split}
\end{equation*}
Furthermore, one can find $\sigma_y^s$ and $\sigma_y^i$ and obtain $\sigma_y^s\otimes \sigma_y^i$
\begin{equation}
\begin{split}
   \sigma_y^s\otimes \sigma_y^i=
   &-(|{s_2,i_1}\rangle\langle {s_1,u}|-|{s_2,u}\rangle\langle {s_1,i_1}|\\&-|{s_1,i_1}\rangle\langle {s_2,u}|+|{s_1,u}\rangle\langle {s_2,i_1})|\\
   =&
   \begin{pmatrix}
   0 & 0 & 1 & 0\\
0 & 0 & 0 & -1\\
1 & 0 & 0 & 0\\
0 & -1 & 0 & 0
   \end{pmatrix}.
\end{split}
\end{equation}
The density operator reads
\begin{widetext}
\begin{equation}
 \rho=|{\psi}\rangle\langle {\psi}|=
    \begin{pmatrix}
   f & e^{i\phi} \sqrt{(1-f)f}\cos{\theta} & e^{i\phi} \sqrt{(1-f)f}\sin{\theta} & 0\\
e^{-i\phi} \sqrt{(1-f)f}\cos{\theta} & (1-f)\cos^2{\theta} & (1-f)\cos{\theta}\sin{\theta} & 0\\
e^{-i\phi} \sqrt{(1-f)f}\sin{\theta} & (1-f)\cos{\theta}\sin{\theta}  & (1-f)\sin^2{\theta} & 0\\
0 & 0 & 0 & 0
   \end{pmatrix}.
\end{equation}
\end{widetext}
The spin-flipped state $\Tilde{\rho}$ can be found in Ref. \citenum{woo} and is written as
\begin{equation}
\begin{aligned}
 \Tilde{\rho}&=(\sigma_y^s\otimes \sigma_y^i)\rho^{*} (\sigma_y^s\otimes \sigma_y^i)\\&=\begin{pmatrix}
  0 & 0 & e^{-i \phi}\sqrt{(1-f)f}\sin{\theta} & 0\\
0 & 0 & 0 & 0\\
e^{i \phi}\sqrt{(1-f)f}\sin{\theta} & 0 & 0 & 0\\
0 & 0 & 0 & 0
   \end{pmatrix},
\end{aligned}
\end{equation}
where $\rho^{*}$ is the complex conjugation of $\rho$. Then it is easy to find the Hermitian matrix
\begin{equation}
\begin{aligned}
 R=&\sqrt{\sqrt{\rho}\Tilde{\rho}\sqrt{\rho}}\\=&\begin{pmatrix}
  0 & 0 & \sqrt{(1-f)f\sin^2{\theta}} & 0\\
0 & 0 & 0 & 0\\
\sqrt{(1-f)f\sin^2{\theta}} & 0 & 0 & 0\\
0 & 0 & 0 & 0
   \end{pmatrix}.
\end{aligned}
\end{equation}

\begin{figure}[htbp]
\centering
\includegraphics[width=.9\linewidth]{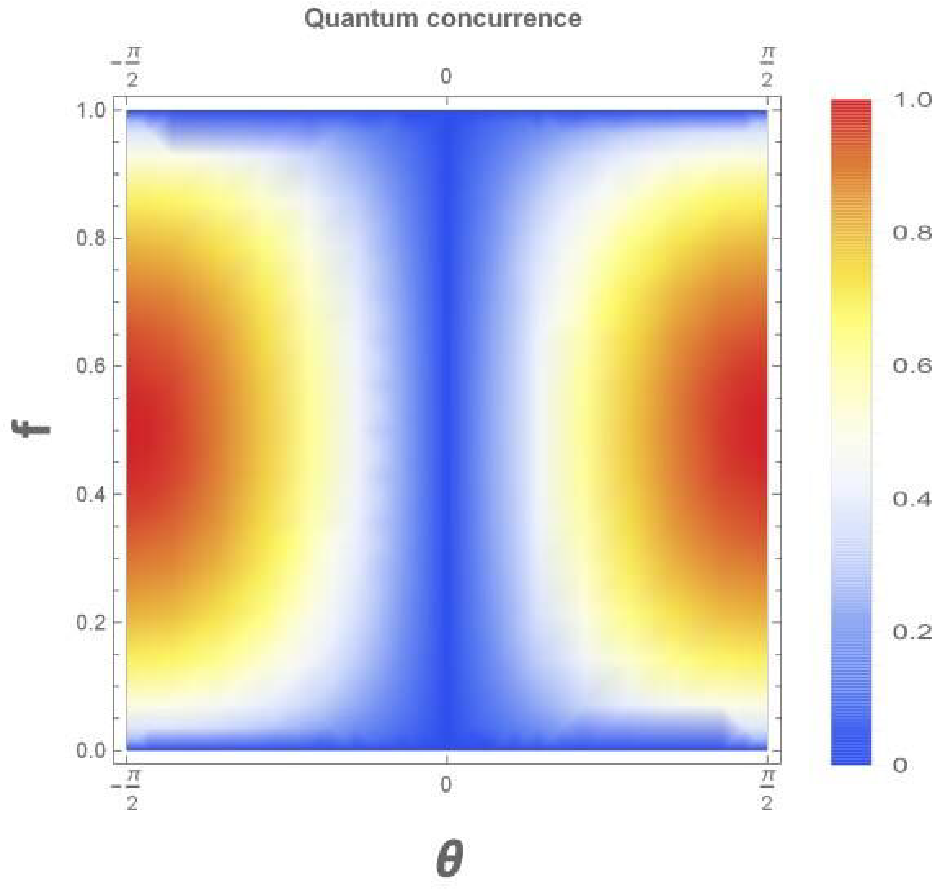}
\caption{Quantum concurrence of the entangled photon pairs state as a function of parameter $\theta$ and $f$.}
\label{fig:false-color}
\end{figure}

Having the eigenvalues of matrix R:  $[\lambda_1,\lambda_2,\lambda_3,\lambda_4]$= $[\sqrt{(1-f)f\sin^2{\theta}},0,0,-\sqrt{(1-f)f\sin^2{\theta}}]$ in decreasing order. The concurrence for a mixed state of two qubits has been derived by Wootters \cite{woo}, can be expressed in terms of
the parameters $\theta$ and $f$
\begin{equation}
\begin{aligned}
C(\theta,f)&=max(0,\lambda_1-\lambda_2-\lambda_3-\lambda_4)\\&=max(0,2\sqrt{(1-f)f}|\sin{\theta}|),
\end{aligned}
\label{a}
\end{equation}
Alternatively, one can also define a non-Hermitian matrix to solve the matrix
\begin{equation*}
 R_a=\rho\Tilde{\rho}=\begin{pmatrix}
  0 & 0 & (1-f)f\sin^2{\theta} & 0\\
0 & 0 & 0 & 0\\
(1-f)f\sin^2{\theta} & 0 & 0 & 0\\
0 & 0 & 0 & 0
   \end{pmatrix}.
\end{equation*}

The real eigenvalues are thus given in decreasing order: $[\alpha_1,\alpha_2,\alpha_3,\alpha_4]=[(1-f)f\sin^2{\theta},0,0,-(1-f)f\sin^2{\theta}]$ and obtain $\lambda_i=\sqrt{\alpha_i}$.  The concurrence can be eventually solved by
\begin{equation}
\begin{aligned}
C(\theta,f)&=max(0,\sqrt{\alpha_1}-\sqrt{\alpha_2}-\sqrt{\alpha_3}-\sqrt{\alpha_4})\\&=max(0,2\sqrt{(1-f)f}|\sin{\theta}|),
\end{aligned}
\label{b}
\end{equation}
which reproduces Eq.(\ref{a}) from Hermitian matrix.

Figure \ref{fig:false-color} plots the result of Eq.(\ref{a}) as well as Eq.(\ref{b}).

\section{Limiting case of $\sigma_f\rightarrow 0$ for the multi-mode description}

In this case, we have
\begin{eqnarray}\label{fi-Phi}
&&f_s(\Omega_1)\Phi(\Omega_1,\Omega_2) \cr && \hskip 0.2in = \exp\bigg(-\frac{\Omega_1^2}{2\sigma_f^2}\bigg)
\exp\bigg[-\frac{(\Omega_1+\Omega_2)^2}{4\sigma_p^2}
-\frac{(\Omega_1-\Omega_2)^2}{2\sigma_0^2}\bigg]\cr && \hskip 0.2in \approx \exp\bigg(-\frac{\Omega_1^2}{2\sigma_f^2}\bigg)
\exp\bigg[-\frac{\Omega_2^2}{2}\Big(\frac{1}{2\sigma_p^2}+\frac{1}{\sigma_0^2}\Big)\bigg]
\end{eqnarray}
if $\sigma_f \ll \sigma_p, \sigma_0$. Then the factorized states for the idler fields are
\begin{eqnarray}
|i_1\rangle &=&\int d\Omega \phi(\Omega)\hat a_i^{\dag}(\Omega)|vac\rangle \cr
 |i_2\rangle &=&\int d\Omega \phi(\Omega)e^{i\Omega\Delta_i}\hat a_i^{\dag}(\Omega)|vac\rangle
\end{eqnarray}
with $\phi(\Omega)\equiv N \exp\bigg[-\frac{\Omega_2^2}{2}\Big(\frac{1}{2\sigma_p^2}+\frac{1}{\sigma_0^2}\Big)\bigg]$. From these two states, we obtain
\begin{eqnarray}
\cos \theta &\equiv &\langle i_1|i_2\rangle = \int d\Omega |\phi(\Omega)|^2 e^{i\Omega\Delta_i}\cr
&=& \exp\bigg(-\Delta_i^2/4\bar T_i^2\bigg)
\end{eqnarray}
with $\bar T_i^2 \equiv \frac{1}{2\sigma_p^2}+\frac{1}{\sigma_0^2}$

\begin{figure*}
\centering
\includegraphics[width=11cm]{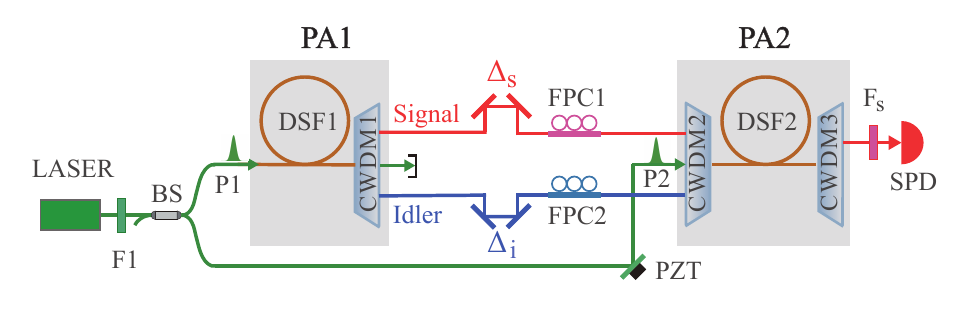}
\caption{Experimental setup. BS, 50/50 beam splitter; CWDM, coarse wavelength division multiplexer; DSF, dispersion shifted fiber; F, filter; P, pump; PA, parametric amplifier; PZT, piezoelectric transducer; SPD, single photon detector.}
\label{setup}
\end{figure*}

\section{Details of the experimental arrangement and procedures}

\subsection{Experimental setup}

The experimental setup of our SU(1,1) interferometer is shown in Fig.~\ref{setup}. The two PAs are based on four-wave
mixing (FWM) in dispersion-shifted fibers (DSF). The pumps of two PAs (P1 and P2) are obtained by passing
the output of a femoto-second fiber laser through a band pass filter (F1) and 50/50 beam splitter (BS). The repetition rate of laser is about 50 MHz. PA1 consists of a DSF and a coarse wavelength division multiplexer (CWDM); while PA2 consists of a DSF and two
CWDMs. The two DSFs with zero-dispersion wavelength and length of 1548.5 nm and 150 m, respectively, are
identical. The central wavelength of the pumps (P1 and P2) is set at 1549.3 nm, and the pulse width (or bandwidth) of two pumps can be varied within the range of about $3$-$15$ ps ($0.2$-$0.8$ nm ) by adjusting F1. Under this condition, the phase matching condition
of FWM with a gain bandwidth up to 25 nm in both signal and idler bands is
achieved in DSFs. The course wavelength division multiplexers
(CWDMs) having three channels for pump, signal,
and idler fields, respectively, are used for the separation
or combination of fields at different wavelengths.

The entangled signal and idler fields are generated in
PA1 via spontaneous FWM. The two entangled fields are
separated from P1 by CWDM1 and propagate along different
paths. CWDM2 couples pump P2 and signal and
idler fields into DSF2, and CWDM3 separates signal and
idler output fields from P2. In the experiment, the production rate of signal and idler photon
pairs is about 0.1 pairs/pulse/nm in PA1. The delays relative to
the time slot of pump pulse ($\Delta_s$ and $\Delta_i$) are introduced
to the signal and idler arms for an unbalanced interferometer.
In PA2, the mode overlap of all fields involved in FWM is optimized by adjusting the fiber polarization controllers (FPC1 and FPC2) in signal and idler arms to maximize the visibility~\cite{Guo-SR16}. When the SU(1,1) interferometer is balanced, i.e., $\Delta_s=\Delta_i=0$, the visibility is the highest. With the increase of the delays, the visibility decreases.

With the intention to lengthen the coherence
time of the detected field and erase the temporal distinguishability
for recovering the interference at the output,
a narrow optical filter (F$_s$) is placed at the signal output
port. The power of filtered signal field is measured by
an InGaAs-based single photon detector (SPD, Langyan
SPD4V5) operated in a gated Geiger mode. The 2.5-ns
gate pulses coincide with the arrival of photons at SPDs.
The response time of SPDs is about 1 ns, which is 100
times longer than the pulse duration of the detected field.
The interference pattern is measured when the phase of
the pump P2 is varied by scanning a piezoelectric transducer
at about 0.8 Hz and the integration time of SPD is
set to 20 ms.

In our experimental setup, the transmission efficiencies for the idler and signal arms between two PA2 are $56\%$ and $64\%$, respectively, because there exist transmission loss of CWDMs, fusion loss between fibers and coupling loss induced by delay lines. Since the transmission loss in each DSF is negligibly small, the ratio between the powers of P2 and P1 is adjusted to $P_2/P_1=0.6$. In this case, time-bin entangled state with maximum entanglement （$A_1=A_2$） can be obtained \cite{cui21-APL22}. For each CWDM, the 1 dB bandwidth of each channel is 16 nm, and the wavelength separation of adjacent channels is 20 nm. So the equivalent 3 dB bandwidth of correlated signal and idler fields launched into DSF2 is about 10 nm (1.25 THz), which is a combined spectral effect of the FWM gain and CWDM pass bands. As a result, the value of $\sigma_0$ in Eqs. (14) should be substituted for this equivalent bandwidth (about 4.8 rad/ps). The bandwidth of the filter (F$_s$) is controlled by using a programmable optical filter (WS, Finisar 4000S), whose central wavelength is fixed at 1566.5 but the 3 dB bandwidth can be flexibly adjusted from 0.02 to 1 THz.

\subsection{Procedure for obtaining normalized visibility}

\begin{figure}
\includegraphics[width=8cm]{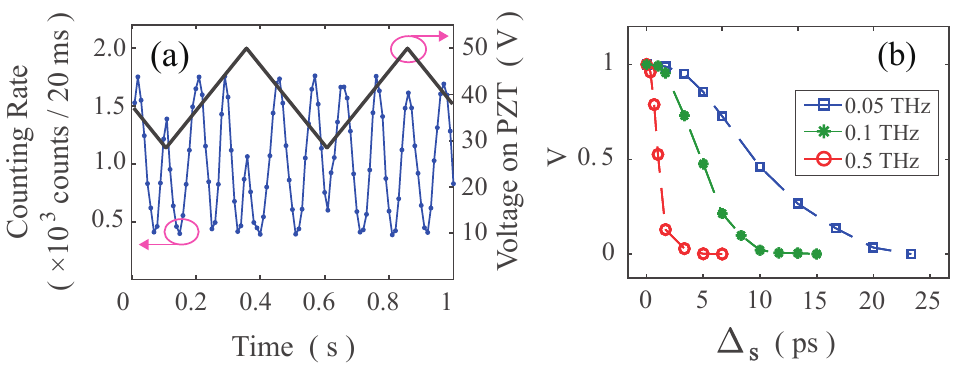}
\caption{(a) The directly observed interference pattern when the phase of the pump P2 is varied by scanning the voltage applied on PZT and $\Delta_s=\Delta_i=0$. (b) The normalized visibility as a function of delay $\Delta_s$ when $\Delta_i=0$ and the 3 dB bandwidth of filter (Fs) is 0.05, 0.1 and 0.5 THz, respectively.}
\label{visibility}
\end{figure}

During the experiment, the 3 dB bandwidth of pump is first set at 0.6 nm. Both the counting rate of SPD and voltage applied on PZT are recorded and analyzed by using a data acquisition system. The blue solid circles in Fig. \ref{visibility}(a) present a typical set of interference pattern when the delays in two arms of the SU(1,1) interferometer are zero ($\Delta_s=\Delta_i=0$) and the phase of the pump P2 is varied by scanning the PZT. Ideally, the visibility should be 100$\%$. However, because of the nonideal transmission between the two PAs and the existence of background Raman scattering in each DSF~\cite{li-opex04}, the visibility in Fig. \ref{visibility}(a) is only about 60$\%$ \cite{cui21-APL22}. With the increase of $\Delta_s$ and $\Delta_i$, the visibility will further decreases.

In principle, by correcting the directly observed interference pattern with the background noise induced by Raman scattering and transmission loss inside the SU(1,1) interferometer \cite{cui21-APL22}, 100$\%$ visibility is obtainable when $\Delta_s=\Delta_i=0$. Considering our goal here is to reveal dependence of time constants (T$_s$ and T$_i$) of the field passing through filter F$_s$, the key is to describe how the visibility varies with the delays $\Delta_s$ and $\Delta_i$. Hence, the loss and Raman scattering induced non-ideal visibility (see Fig. \ref{visibility}(a)) will not affect the evaluation of T$_s$ and T$_i$. This is similar to the technique in characterizing the coherence time of an optical field by using a Mach-Zehnder interferometer, whose maximum visibility deviates from 100$\%$ due to the unequal intensities in two arms.

In the experiment, we record the interference patterns for $\Delta_s$ and $\Delta_i$ with different values. Instead of carefully characterizing the amount of background noise to correct the raw data of interference, we normalize the observed visibility V$_o$ with V$_{max}$, which is directly observed under the condition of $\Delta_s=\Delta_i=0$. As an example, Fig. \ref{visibility}(b) presents the normalized visibility ($V=V_o/V_{max}$) as a function of delay $\Delta_s$ when the delay $\Delta_i$ is fixed at 0 and the 3 dB bandwidth of filter F$_s$ is 0.05, 0.1, and 0.5 THz, respectively. By collecting the normalized visibility for F$_s$ with different bandwidths when the delays are $\Delta_i=0$, $\Delta_s=10$ ps, trace (i) (blue squares) in Fig. 1 of the main text is then acquired. The other two traces in Fig. 1 are obtained by collecting the normalized visibility obtained under the conditions of $\Delta_s=10$, $\Delta_i=5$ ps and $\Delta_s=\Delta_i=10$ ps, respectively.

\subsection{Procedure for deducing time constants T$_s$ and T$_i$}

\begin{figure}
\includegraphics[width=8cm]{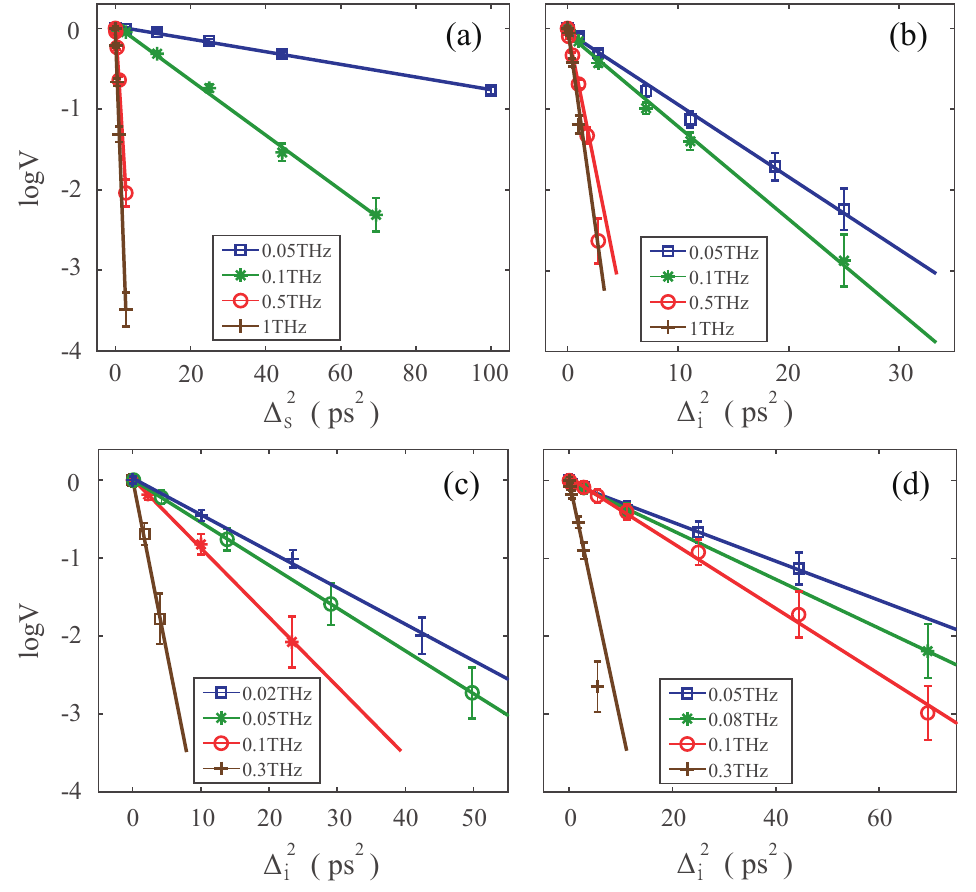}
\caption{The normalized visibility as a function of (a) $\Delta_s^2$  with $\Delta_i=0$ and (b,c,d) $\Delta_i^2$ with $\Delta_s=0$ for F$_s$ having different bandwidths as indicated in legends when the 3 dB bandwidth of pump is 0.24, 0.6 and 0.8 nm, respectively.}
\label{time-constant}
\end{figure}

From the normalized visibility obtained under the condition of $\Delta_i=0$ (see Fig. \ref{visibility}(b)), we extract the time constant T$_s$ by using Eq. (15) and fitting data to a linear regression of $\log V$ vs. $\Delta_s^2$, as shown in Fig. \ref{time-constant}(a). Similarly, by using Eq. (18), we can extract the time constant T$_i$ by changing $\Delta_i$ and fixing $\Delta_s$ at 0 when bandwidth of the filter F$_s$ takes different values, as shown in Fig. \ref{time-constant}(c). Moreover, we repeat the measurement of T$_s$ and T$_i$ by changing the 3 dB bandwidth of pump to 0.24 nm and 0.8 nm, respectively. We find there is no observable change in the deduced value of T$_s$ when the pump bandwidth is altered. However, as shown in Fig. \ref{time-constant}(b) and (d),  T$_i$ varies with the pump bandwidth. According to the best fit in Fig. \ref{time-constant}, the four traces in Fig. 3 of the main text are obtained.